\documentclass[pra,aps,preprint]{revtex4-1}
\usepackage{amsmath}
\usepackage{amssymb}
\usepackage{graphicx}

\newcommand{\mP}{{\mathcal P}}
\newcommand{\mT}{{\mathcal T}}

\newcommand{\beqa}{\begin{eqnarray}}
\newcommand{\eeqa}{\end{eqnarray}}

\begin{document}
\title{Robust and fragile $\mP\mT$-symmetric phases in a tight-binding chain} 
\author{Yogesh N. Joglekar}
\email{yojoglek@iupui.edu}
\author{Derek Scott}
\author{Mark Babbey}
\affiliation{Department of Physics, 
Indiana University Purdue University Indianapolis (IUPUI), 
Indianapolis, Indiana 46202, USA}
\author{Avadh Saxena}
\affiliation{Theoretical Division, Los Alamos National Laboratory, Los Alamos, New Mexico 87545, USA}
\date{\today}
\begin{abstract}
We study the phase-diagram of a parity and time-reversal ($\mP\mT$) symmetric tight-binding chain with $N$ sites and hopping energy $J$, in the presence of two impurities with imaginary potentials $\pm i\gamma$ located at arbitrary ($\mP$-symmetric) positions $(m, \bar{m}=N+1-m)$ on the chain where $m\leq N/2$. We find that except in the two special cases where impurities are either the farthest or the closest, the $\mP\mT$-symmetric region - defined as the region in which all energy eigenvalues are real - is algebraically fragile. We analytically and numerically obtain the critical impurity potential $\gamma_{PT}$  and show that $\gamma_{PT}\propto 1/N\rightarrow 0$ as $N\rightarrow\infty$ except in the two special cases. When the $\mP\mT$ symmetry is spontaneously broken, we find that the maximum number of complex eigenvalues is given by $2m$. When the two impurities are the closest, we show that the critical impurity strength $\gamma_{PT}$ in the limit $N\rightarrow\infty$ approaches $J$ ($J/2$) provided that $N$ is even (odd). For an even $N$ the $\mP\mT$ symmetry is maximally broken whereas for an odd $N$, it is sequentially broken. Our results show that the phase-diagram of a $\mP\mT$-symmetric tight-binding chain is extremely rich and that, in the continuum limit, this model may give rise to new $\mP\mT$-symmetric Hamiltonians. 
\end{abstract}
\maketitle

\noindent{\it Introduction:} Lattice models, including tight-binding chains, have been a cornerstone of theoretical explorations due to their analytical and numerical tractability~\cite{wen}, the absence of divergences associated with the ultraviolet cutoff~\cite{kogut1,kogut2}, the availability of exact solutions~\cite{onsager}, and the ability to capture counter-intuitive physical phenomena including the bound-states in repulsive potentials~\cite{winkler}. In recent years, sophisticated optical lattice systems have increasingly permitted the experimental exploration of lattice models~\cite{bloch,zoller}. These lattice models are largely based on a Hermitian Hamiltonian. Over the past decade, it has become clear that non-Hermitian Hamiltonians with $\mP\mT$-symmetry can have purely real energy spectra~\cite{bender,bender2} and that, when they do, with an appropriately re-defined inner product, their eigenvectors can be appropriately orthonormalized~\cite{bender3}. The theoretical work has been accompanied, most recently, by experiments in optical physics where spontaneous $\mP\mT$ symmetry breaking in a classical system has been observed in waveguides with $\mP\mT$ symmetric complex refractive index~\cite{ruter}. In recent years, lattice models with a $\mP\mT$ symmetric {\it non-Hermitian} ``hopping'' between adjacent levels of a simple-Harmonic oscillator~\cite{znojilco}, tridiagonal $\mP\mT$-symmetric models~\cite{znojiltri}, and tight-binding models with a Hermitian hopping and $\mP\mT$-symmetric complex on-site potential~\cite{bendix,song,song2} have been investigated. These models are of physical interest because they lead to unitary scattering even in the absence of a Hermitian Hamiltonian~\cite{znojilsc}. 

In this paper, we analytically and numerically investigate the phase-diagram of a one-dimensional tight-binding chain with hopping energy $J$ and two imaginary potentials $\pm i\gamma$ as a function of the size of the chain $N$ and the positions of the two impurities within the chain $(m,\bar{m})$~\cite{song2,znojilsc}. Our main results are as follows: i) We show that except in the two special cases (the impurities are the closest or the farthest), in the thermodynamic limit $N\rightarrow\infty$, the critical potential strength vanishes, $\gamma_{PT}(\mu)\propto 1/N\rightarrow 0$ where $\mu=m/N\leq 0.5$ is the relative position of the impurity. Thus, for a generic impurity location, the $\mP\mT$-symmetric phase in this system is algebraically fragile~\cite{nb}. ii) The ``degree of $\mP\mT$ symmetry breaking'', defined as the fraction of eigenvalues that become complex, is given by $2\mu$. iii) When the impurities are the closest, the critical potential strength is given by $\gamma_{PT}(\mu=0.5)=J$ when $N$ is even and when $N$ is odd, $\gamma_{PT}(\mu\sim 0.5)\rightarrow J/2$ as $N\rightarrow\infty$. iv) By considering the continuum limit of this problem, we argue that the $\mP\mT$-symmetric chain maps onto a $\mP\mT$-symmetric continuum Hamiltonian with an imaginary viscous drag term. 

\noindent{\it Tight-binding Model:} 
We start with the Hamiltonian for an $N$-site tight-binding chain with $J>0$ and two impurities with imaginary potentials at sites $(m,\bar{m}=N+1-m)$,  
\begin{equation}
\label{eq:h0}
\hat{H}=-J\sum_{n=1}^{N}\left( c^{\dagger}_{n}c_{n+1} + c^{\dagger}_{n+1}c_n\right)+i\gamma\left(c^{\dagger}_m c_m-c^{\dagger}_{\bar{m}}c_{\bar{m}}\right),
\end{equation}
where $c^{\dagger}_n$ ($c_n$) is the creation (annihilation) operator at site $n$, $m\leq N/2$ and $\bar{m}=(N+1-m)> N/2$ is the reflection-counterpart of site $m$. We only consider the single-particle sector and, since periodic-boundary conditions are incompatible with the parity operator, our system has open boundary conditions. In the lattice model, the action of the parity operator is given by $\mP c^{\dagger}_n\mP=c^{\dagger}_{\bar{n}}$ with $\bar{n}=(N+1-n)$, and the action of the anti-linear time-reversal operator is given by $\mT i\mT=-i$. The potential term in the Hamiltonian, Eq.(\ref{eq:h0}),  is odd under individual $\mP$ and $\mT$ operations, and hence the Hamiltonian is $\mP\mT$-symmetric. The phase-diagram of this system with the impurities {\it at the end} and its equivalent Hermitian counterpart have been investigated in detail~\cite{song}. 

A general single-particle eigenfunction of the Hamiltonian can be written as $|\psi_k\rangle=\sum_{n=1}^{N}\psi^{k}_n c^{\dagger}_n |0\rangle=\sum_n\psi^{k}_n |n\rangle$ where, according to the Bethe ansatz, the coefficients $\psi^k_n$ have the form
\begin{equation}
\label{eq:psikn}
\psi^{k}_n=\left\{\begin{array}{cc}
A\sin(kn) & n\leq m,\\
P\sin(kn)+ Q\cos(kn) & m< n<\bar{m},\\
B\sin(k\bar{n}) & n\geq \bar{m}.\\
\end{array}\right.
\end{equation}
Note that open-boundary conditions, along with the requirement $\hat{H}|\psi_k\rangle=E_k|\psi_k\rangle$, constrain the eigenfunction coefficients to a sinusoidal form in the regions $n\leq m$ and $n\geq\bar{m}$ and give $E_k=-2J\cos(k)$. By considering the eigenvalue equation at points $n=m, m+1$ and their reflection-counterparts, we find that the quasimomenta $k$ obey the equation
\begin{eqnarray}
\label{eq:quasik}
M(k) & \equiv & \left[\sin^2(k(m+1))+\left(\frac{\gamma^2}{J^2}\right)\sin^2(km)\right]\sin(k(N-2m+1))\nonumber\\
& + & \sin^2(km)\sin(k(N-2m-1)) -2\sin(km)\sin(k(m+1))\sin(k(N-2m))= 0.
\end{eqnarray}
The $\mP\mT$ symmetry is unbroken provided that Eq.(\ref{eq:quasik}) has $N$ real solutions. When $\gamma=0$, the $N$ distinct solutions are given by the well-known result for a tight-binding chain with open boundary conditions, $k_\alpha=\alpha\pi/(N+1)$ where $\alpha=1,\ldots,N$. Since $M(\pi-k)=(-1)^N M(k)$, if $k_0$ is a solution of Eq.(\ref{eq:quasik}), then so is $(\pi-k_0)$.  It also follows that when $N$ is odd, $k=\pi/2$ is a solution irrespective of the value of $\gamma$, and that the corresponding eigenvector has zero energy. When $m=1$, Eq.(\ref{eq:quasik}) reduces to the criterion for quasimomentum obtained in Ref.~\cite{song}; in that case, as $\gamma/J\rightarrow 1$, the two central $k_\alpha\sim \pi/2$ become degenerate, the $\mP\mT$ symmetry is spontaneously broken, and the system develops $N-2$ real and two complex (conjugate) eigenvalues.

\begin{figure}[h!]
\begin{center}
\begin{minipage}{20cm}
\begin{minipage}{9cm}
\hspace{-3cm}
\includegraphics[angle=0,width=9cm]{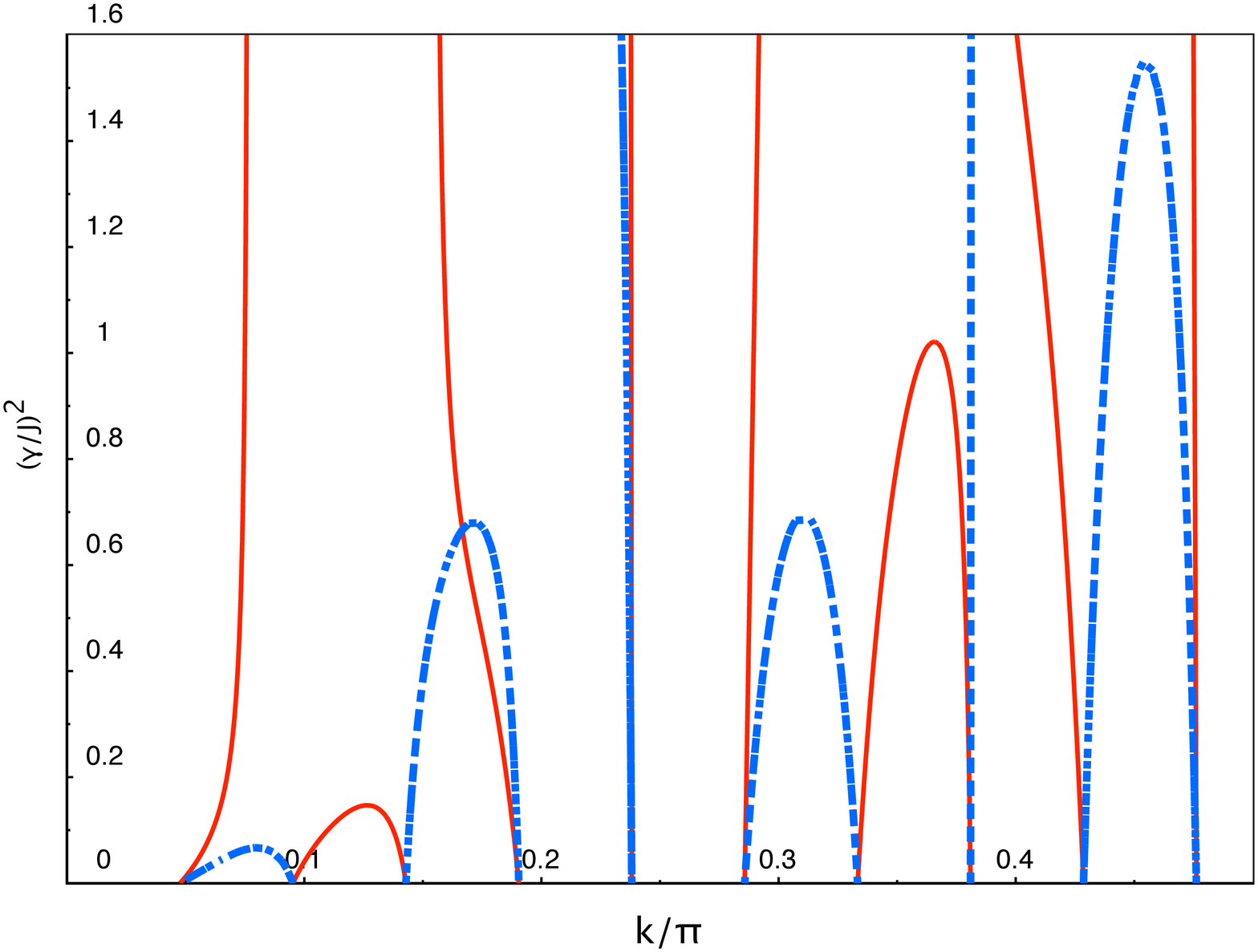}
\end{minipage}
\begin{minipage}{9cm}
\hspace{-4cm}
\includegraphics[angle=0,width=9cm]{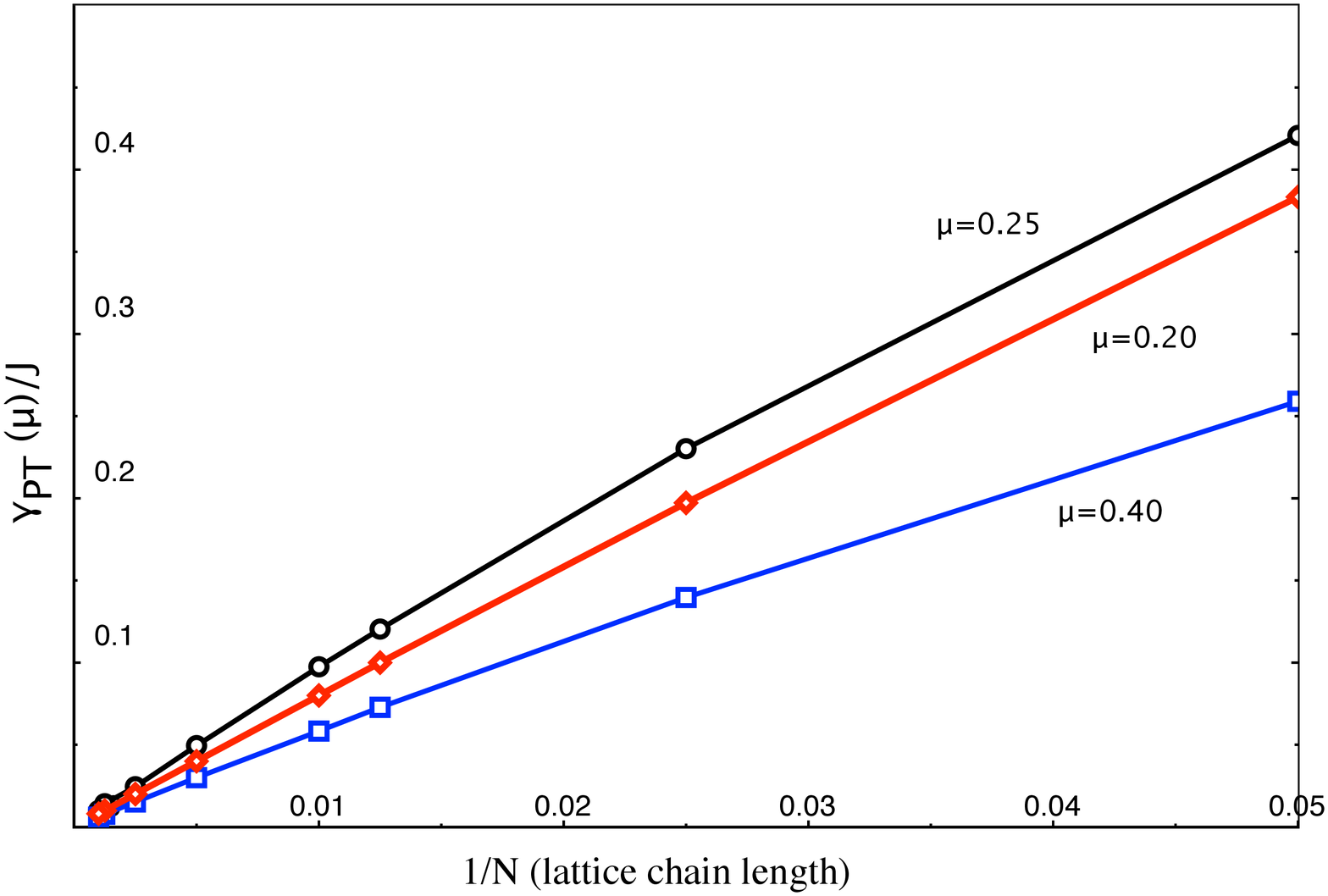}
\end{minipage}
\end{minipage}
\caption{\label{fig:ptphase}
(color online) (a) Left panel shows the permitted values of quasimomenta $k/\pi$ as a function of impurity potential $(\gamma/J)^2$ for a chain with $N=20$ sites and one impurity at $m=4$ ($\mu=0.2$, thin solid red) or at $m=8 $ ($\mu=0.4$, thick dashed blue). The plot is symmetric about $k=\pi/2$. As $(\gamma/J)^2$ is increased, the adjacent quasimomenta $k_1\sim \pi/(N+1)$ and $k_2\sim 2\pi/(N+1)$ become degenerate (as do their counterparts near $k\sim\pi$) and lead to the $\mP\mT$ symmetry breaking. (b) Right panel shows the critical potential strength $\gamma_{PT}(\mu)/J$ as a function of the chain size $N$ for various positions $m=\mu N$ of the impurity potential. We see that $\gamma_{PT}(\mu)$ vanishes as $N\rightarrow \infty$; thus the $\mP\mT$-symmetric phase is algebraically fragile except when the impurities are either closest to each other or the farthest.}
\end{center}
\end{figure} 
The left panel in Fig.~\ref{fig:ptphase} shows the typical plot of quasimomentum values $k(\gamma)/\pi$ as a function of $\gamma$, for a chain with $N=20$ sites and the first impurity at  $m$=4 (thin solid red) or $m$= 8 (thick dashed blue). Since the plot is symmetric about $k=\pi/2$, we focus only on the left-half $k/\pi\leq 0.5$ and note that since $N$ is even, there is no solution present at $k=\pi/2$. As the impurity potential $\gamma$ is increased, the adjacent quasimomenta and the corresponding energy levels become degenerate, leading to the $\mP\mT$ symmetry breaking~\cite{bendix,song}. We see that the critical potential for $m=4$ is greater than that for $m=8$, $\gamma_{PT}(\mu=0.2)> \gamma_{PT}(\mu=0.4)$ and, in contrast with the $m=1$ case, the two levels that become degenerate occur in a pair, with one near the origin $k\sim \pi/(N+1)$ and other near the zone-boundary $k\sim N\pi/(N+1)$. Therefore when $\gamma(\mu)=\gamma_{PT}(\mu)+0^{+}$, there are four complex eigenvalues. Since the $\mP\mT$ symmetry breaking can be  associated with the development of dissipative channels,  we define the ``degree of $\mP\mT$-symmetry breaking'' as the fraction of eigenvalues that become complex. For a general $m$, as $\gamma$ is increased beyond $\gamma_{PT}$, we find that a total of $2m$ complex eigenvalues emerge and thus the degree of $\mP\mT$-symmetry breaking is given by $2\mu\leq 1$. The right panel in Fig.~\ref{fig:ptphase} shows the typical evolution of critical potential strength $\gamma_{PT}(\mu)$ with $N$, for $\mu=0.2, 0.25, 0.4$, obtained by numerically solving Eq.(\ref{eq:quasik}). The scaling suggests that the critical potential strength for the infinite chain approaches zero, $\gamma_{PT}(\mu)\propto 1/N\rightarrow 0$. {\it Thus, the $\mP\mT$-symmetric phase, which exists in the region $0\leq \gamma\leq\gamma_{PT}$, is algebraically fragile.} This result can be qualitatively understood as follows: in the limit $N\gg 1$ and $m\gg 1$ with $m/N=\mu$, Eq.(\ref{eq:quasik}) can be approximated by $\gamma^2\sin\left[(1-2\mu)kN\right]=0$. If $\gamma\neq 0$, this equation has only $N(1-2\mu)<N$ real solutions, and hence the $\mP\mT$ symmetry is broken. We emphasize that this argument is invalid when $\mu\rightarrow 0$ and the corresponding critical potential is given by $\gamma_{PT}(\mu)=J$~\cite{song}. It is also invalid when $\mu\rightarrow 1/2$, the impurities are closest to each other and, as we discuss below, the critical $\gamma_{PT}(\mu)$ is nonzero when $N\rightarrow\infty$. Incidentally, we point out that the corresponding $\mP\mT$-symmetric phase in a tridiagonal Hamiltonian with {\it non-Hermitian} ``hopping'' is not algebraically fragile~\cite{znojiltri}.  

\begin{figure}[h!]
\begin{center}
\includegraphics[angle=0,width=9cm]{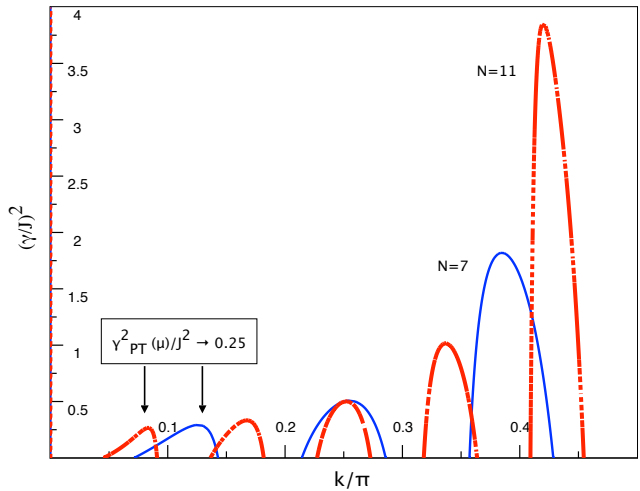}
\caption{\label{fig:nnn}
(color online) Permitted quasimomenta $k(\gamma)/\pi$ for a chain with $N=21$ (thick dotted red) and $N=13$ (thin solid blue) sites as a function of impurity strength $(\gamma/J)^2$ when the impurities are closest to each other. As $\gamma$ is increased, adjacent quasimomenta become degenerate leading to a spontaneous $\mP\mT$-symmetry breaking. As $N\rightarrow\infty$, we find that the critical potential strength $\gamma_{PT}\rightarrow J/2$.} 
\end{center}
\end{figure}

\noindent{\it Closest Impurities and the Even-Odd Effect:} We now consider the case of closest impurities. Note that due to the $\mP\mT$-symmetric requirement, when $N$ is even the impurities are nearest neighbors with $m=N/2$, whereas when $N$ is odd the impurities are next-nearest-neighbors with $m=(N-1)/2$. We will first focus on the case with an even $N$. In this case, the condition $M(k)=0$ from Eq. (\ref{eq:quasik}) reduces to the following equation
\begin{equation}
\label{eq:nn}
J^2\sin^2\left[k\left(\frac{N}{2}+1\right)\right]=(J^2-\gamma^2)\sin^2\left(\frac{kN}{2}\right).
\end{equation}
When $\gamma=0$ Eq.(\ref{eq:nn}) has $N$ distinct solutions given by $k_\alpha=\alpha\pi/(N+1)$. As $\gamma$ increases the adjacent $k_\alpha$ approach each other and when $\gamma=J$, Eq.(\ref{eq:nn}) has $N/2$ doubly-degenerate solutions given by $k_n=2n\pi/(N+2)$ where $n=1,\ldots,(N/2)$. When $\gamma> J$, it is clear that Eq.(\ref{eq:nn}) has no real solutions. {\it Thus the $\mP\mT$ symmetry is maximally broken and all $N$ eigenvalues simultaneously become complex}. When $N$ is odd, the impurities are at sites $N_0\pm 1$ where 
$N_0=(N+1)/2$ is the site at the center of the chain. The equation $M(k)=0$ then reduces to 
\begin{equation}
\label{eq:nnn}
\cos(k)\left[\sin^2(kN_0)+\left(\frac{\gamma^2}{J^2}\right)\sin^2(k(N_0-1))\right]=\sin(kN_0)\sin(k(N_0-1)).
\end{equation}
This equation has all real solutions provided $\gamma/J\leq 1/(2\cos k_d)$ where $\pi/(N+1)< k_d< 2\pi/(N+1)$ corresponds to the first degenerate quasimomentum. Therefore, we find that as $\gamma$ is increased from zero the $\mP\mT$-symmetry breaks at $\gamma_{PT}=J/2$ in the limit $N\rightarrow\infty$ when adjacent $k_\alpha$ near the origin (and their counterparts near the zone boundary) become degenerate. Hence, for $\gamma=\gamma_{PT}+0^{+}$, there are four complex eigenvalues. On the other hand, Eq.(\ref{eq:nnn}) has only one real solution, $k=\pi/2$, when $\gamma/J> 1/(2\cos k_D)$ where $k_D\lesssim \pi/2$ is the degenerate quasimomentum closest to the zone center $k=\pi/2$. Hence, the number of complex eigenvalues increases monotonically and when $\gamma> J/(2\cos k_D)\sim J(N+1)/3\pi$, it saturates to $2m=N-1$~\cite{caveat}. Figure~\ref{fig:nnn} shows the quasimomenta $k_\alpha(\gamma)$ for a chain with $N=13$ and $N=21$ lattice sites obtained from Eq.(\ref{eq:nnn}), and confirms these results. Since the $\mP\mT$ symmetric nature  of the potential dictates the minimum distance between the impurities when $N$ is odd or even, the phase-diagram of the chain is sensitive to it even as $N\rightarrow\infty$. 

\begin{figure}[h!]
\begin{center}
\begin{minipage}{20cm}
\begin{minipage}{9cm}
\hspace{-3cm}
\includegraphics[angle=0,width=8cm]{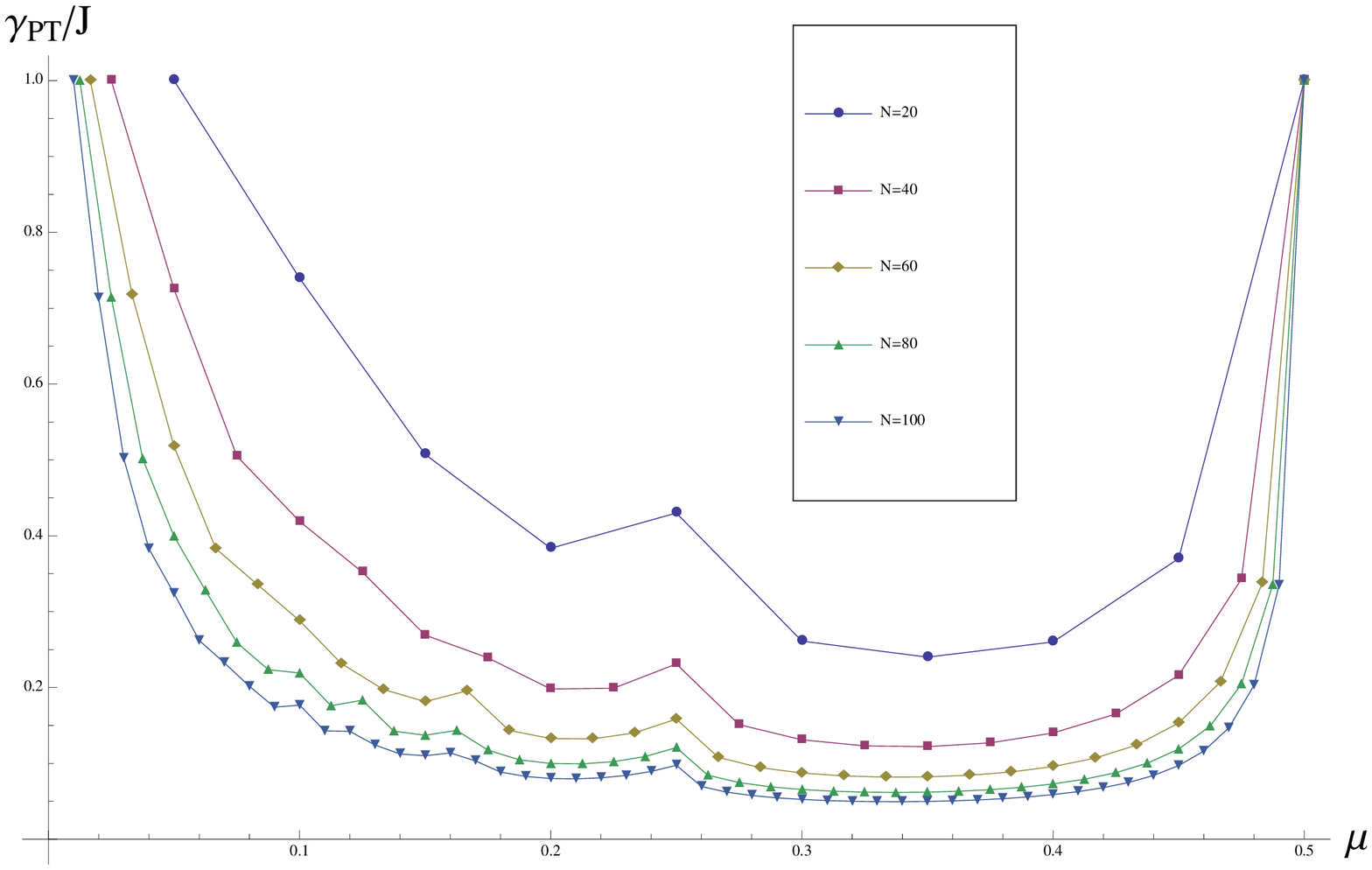}
\end{minipage}
\begin{minipage}{9cm}
\hspace{-4cm}
\includegraphics[angle=0,width=8cm]{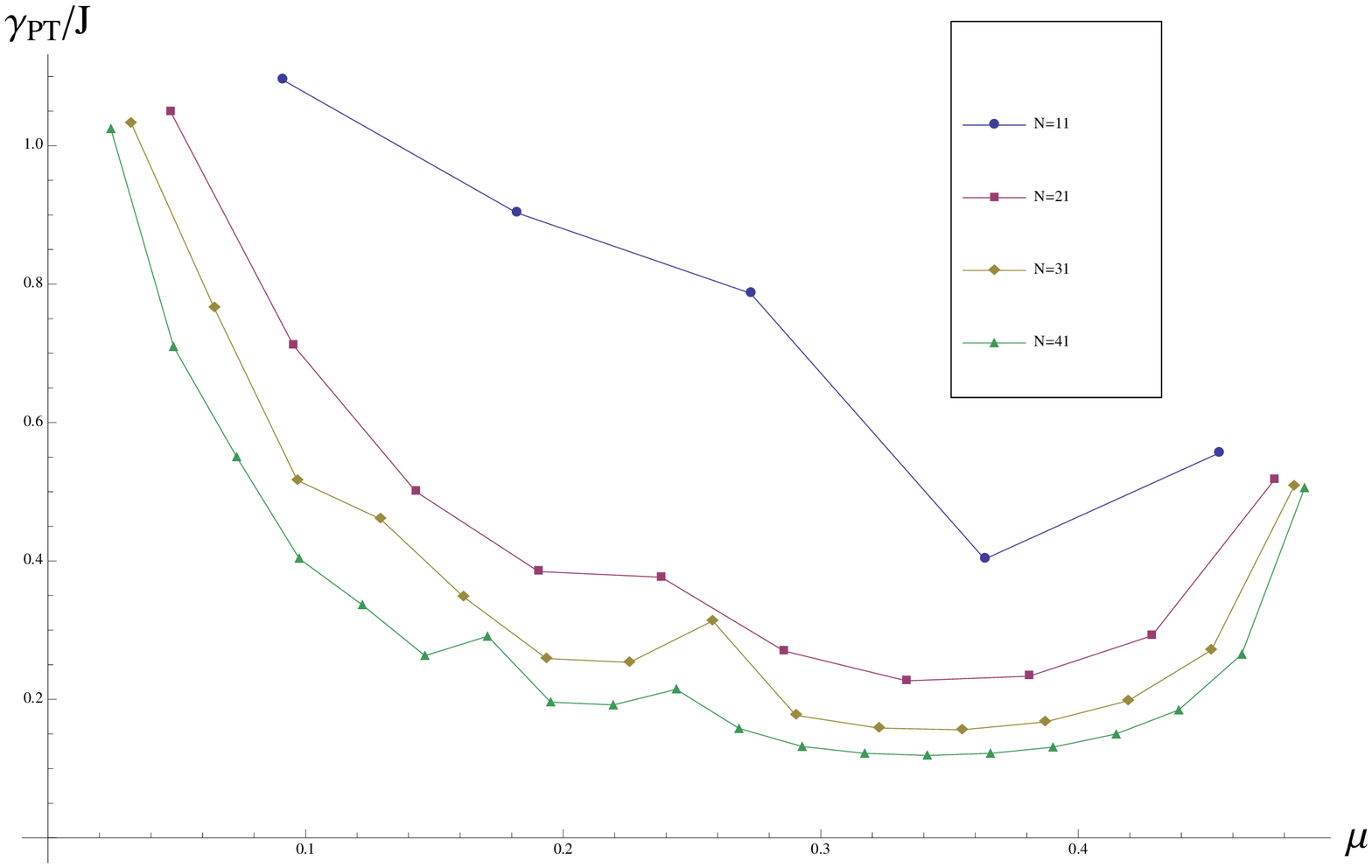}
\end{minipage}
\end{minipage}
\caption{\label{fig:chain}
(color online) (a) Left panel shows the critical potential strength $\gamma_{PT}/J$ as a function of the location $\mu=m/N\leq 1/2$ of the first impurity obtained via numerical diagonalization for a chain with even $N$. Except at the end-points, $\mu=1/N$ and $\mu=0.5$, as $N$ increases $\gamma_{PT}(\mu)$ decreases. At the end points, we find that $\gamma_{PT}=J$ is independent of the value of $N$. (b) The right panel presents similar results for an odd $N$, showing that the critical $\gamma_{PT}(\mu)\propto 1/N$ for all values of $\mu$ except for $\mu=1/N$ and $\mu=0.5-1/2N$.  When $\mu=1/N$, we recover the result $\gamma_{PT}=J\sqrt{1+1/N}$~\cite{song}, and when the impurities are next-nearest-neighbors, we find that the critical potential strength $\gamma_{PT}\rightarrow J/2$ as $N\rightarrow\infty$.}
\end{center}
\end{figure}

\noindent{\it Numerical Results:}  We start this section with results for the critical potential strength $\gamma_{PT}(\mu)$ as a function of the relative impurity site location $0<\mu=m/N\leq 1/2$ obtained by numerical diagonalization of the Hamiltonian Eq.(\ref{eq:h0}) for various lattice sizes $N$, even and odd. The left panel in Fig.~\ref{fig:chain} shows that, for an even $N$, apart from finite-size effects that are prominent near $\mu=1/4$ and are also present in solutions to Eq. (\ref{eq:quasik}), the critical potential strength $\gamma_{PT}(\mu)$ is vanishingly small except at $\mu=1/N$ and $\mu=0.5$. In both special cases $\gamma_{PT}=J$. The right panel in the same figure shows results for odd $N$. When $\mu=1/N$ or equivalently $m=1$, we recover the result $\gamma_{PT}=J\sqrt{1+1/N}$~\cite{song}. As in the case with even $N$, we find that $\gamma_{PT}(\mu)$ is suppressed with increasing $N$ everywhere except when $\mu=0.5-1/2N$ or equivalently $m=(N-1)/2$. These results are consistent with those obtained through the analytical treatment earlier. 

\begin{figure}[h!]
\begin{center}
\includegraphics[angle=0,width=9cm]{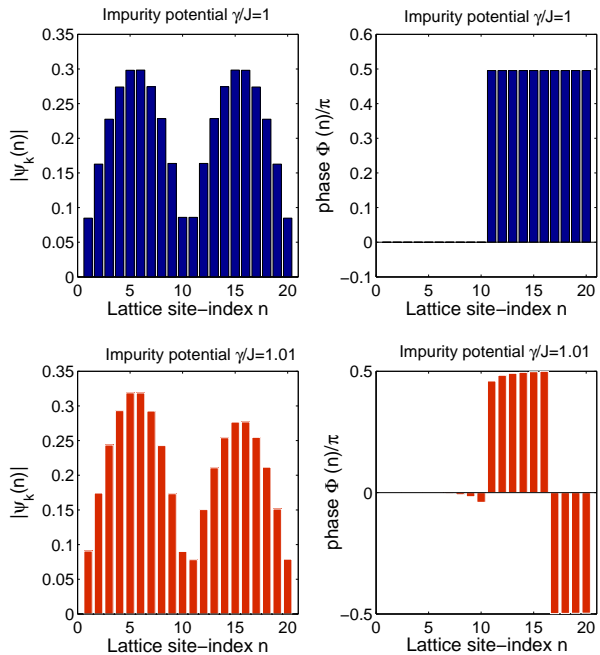}
\caption{\label{fig:wavefunction}
(color online) The top (blue) panel shows the amplitude $|\psi_k(n)|$ (left) and the phase $\Phi(n)/\pi$ (right) of the ground-state wavefunction of a $\mP\mT$-symmetric chain with $N=20$ sites and nearest-neighbor impurities with strength $\gamma=\gamma_{PT}=J$. As expected of a $\mP\mT$-symmetric state, the amplitude is even around the center of the chain, and the effect of a non-zero $\gamma$ is manifest in the discontinuous change in the phase, with $\Phi(n)=0$ for $n\leq N/2$ and $\Phi(n)=\pi/2$ for $n>N/2$, consistent with Eq.(\ref{eq:wavefunction}). The bottom (red) panel shows the same state when $\gamma/J=1.01$ and the $\mP\mT$-symmetry is spontaneously broken. The broken symmetry is manifest in the asymmetrical wavefunction amplitude (left) and a position-dependent phase $\Phi(n)$.}
\end{center}
\end{figure}

We now briefly explore the change in a (typical) eigenfunction $\psi_k(n)$ as a function of impurity potential $\gamma$ in the case of nearest-neighbor impurities (even $N$). In the $\mP\mT$-symmetric phase, an eigenfunction is given by $\psi_<(n)=A\sin(kn)$ for $n\leq N/2$ and $\psi_>(n)=B\sin(k\bar{n})$ for $n> N/2$ where $k$ is a quasimomentum that satisfies Eq.(\ref{eq:nn}). Using the eigenfunction constraints and Eq.(\ref{eq:nn}), it follows that 
\begin{equation}
\label{eq:wavefunction}
B= A\left[\frac{\sin(k(1+N/2))}{\sin(kN/2)}+i\frac{\gamma}{J}\right]= A\exp\left(i\theta_{\gamma}\right),
\end{equation}
where the angle $\theta_\gamma$ satisfies $\tan\theta_\gamma=\gamma\sin(kN/2)/J\sin\left[k(1+N/2)\right]$. Figure~\ref{fig:wavefunction} shows the amplitude $|\psi_k(n)|$ and the phase $\Phi(n)$ of the ground-state wavefunction of a chain with $N=20$ sites and nearest-neighbor impurities. The top (blue) panel shows that when $\gamma=\gamma_{PT}=J$, the wavefunction amplitude is even about the center of the chain and the phase is given by $\theta_\gamma=\pi/2$, as is expected from Eq.(\ref{eq:wavefunction}). The bottom (red) panel shows that when $\gamma=1.01 J>\gamma_{PT}/J$, {\it the broken $\mP\mT$-symmetry is reflected in the asymmetrical wavefunction amplitudes and in the position-dependent phase factor $\Phi(n)$}. These are generic features of the broken $\mP\mT$-symmetry phase. We also note that in the continuum limit, the eigenfunction $\psi_k(x)$ becomes {\it discontinuous} at the center of the chain while the probability amplitude $|\psi_k(x)|$ remains continuous. 

\noindent{\it Conclusion:} We have investigated the phase diagram of an $N$-site one-dimensional chain with a pair of complex $\mP\mT$-symmetric impurities located at sites $(m,\bar{m})$ within it. A remarkable feature of such a Hamiltonian, in contrast to a tridiagonal Hamiltonian with real entries~\cite{znojilco}, is that in the $\mP\mT$-symmetric region, its spectrum remains confined within the energy band $\pm 2J$ of the model in the absence of impurities; as the impurity potential $\gamma$ is increased, the level spacing between adjacent energy levels decreases. Our results show that the $\mP\mT$-symmetric phase of such a chain is algebraically fragile except when the impurities are farthest from each other or are closest to each other. In the latter case, we find that the $\mP\mT$-symmetric phase survives when $\gamma\leq \gamma_{PT}=J$ (even $N$) or $\gamma\leq\gamma_{PT}=J/2$ (odd $N$).  We note that such a chain offers tremendous tunability due to its variable critical impurity strength $\gamma_{PT}(\mu)$ for a finite $N$, and the corresponding variable fraction of complex eigenvalues $2\mu$ which translates into the number of dissipative channels in both classical~\cite{ruter} and quantum systems.  Thus, a physical realization of such a model~\cite{song2} may offer the ability to engineer the level-spacings and the dissipation in this system. 

We conclude by briefly pointing out the continuum limit of this system. In the continuum limit, the lattice spacing $a$ vanishes and the number of lattice sites $N$ diverges such that the length of the chain $L=Na$ remains constant. Note that since $\sum_j (V_{PT}\psi)(j)$ is a constant, where $V_{PT}$ is the $\mP\mT$-symmetric potential, the corresponding continuum Schr\"{o}dinger eigenvalue equation is given by 
\begin{equation}
\label{eq:cont}
-\frac{d^2}{dx^2}\psi_k(x)-i\Gamma\delta(x)\frac{d}{dx}\psi_k(x)=k^2\psi_k(x),
\end{equation}
where the dimensionless impurity strength $\Gamma=\gamma/J$ for nearest-neighbor impurities (even $N$), $\Gamma=2\gamma/J$ for next-nearest-neighbor impurities (odd $N$), and the eigenfunctions obey boundary conditions $\psi_k(x=\pm L/2)=0$. Note that the continuum Hamiltonian is $\mP\mT$-symmetric, but not Hermitian, due to the non-commuting parts, $\delta(x)$ and $-id/dx$, of the ``potential" term. Our results imply that the $\mP\mT$-symmetric phase of this Hamiltonian survives as long as $\Gamma\leq 1$, irrespective of whether the number of sites $N$ in the chain is odd or even. Indeed, Eq.(\ref{eq:cont}) suggests a new class of $\mP\mT$-symmetric Hamiltonians with a ``viscous drag potential" term of the form $V=-i\hbar f(x) d/dx$ that is not Hermitian but is $\mP\mT$-symmetric provided $f(x)$ is an even function of $x$. In the lattice model, such a potential will correspond to $\mP\mT$-symmetric impurities at multiple locations. Detailed investigation of such models will improve our understanding of $\mP\mT$-symmetry breaking in discrete and (classical) continuum systems, that can be realized in optical lattices and waveguides with complex refractive index, respectively~\cite{ruter}. 

Y.J. is thankful to Los Alamos National Laboratory where part of this work was carried out. M.B. thanks the D.J. Angus Scientech Foundation for a Summer Fellowship.




\begin{thebibliography}{99}
\bibitem{nb} We use the term ``robust'' $\mP\mT$-symmetric phase provided that $\lim_{N\rightarrow\infty}
\gamma_{PT}(N)>0$. This usage is different from that in Ref.~\cite{znojilco}, but is consistent with 
that in Ref.~\cite{bendix}.
\bibitem{caveat} This is strictly true only when $N=4p+1$. For $N=4p-1$, as $\gamma$ increases, 
the number of real eigenvalues reduces from $7\rightarrow 3\rightarrow 5\rightarrow 1$.
\end{thebibliography}
\end{document}